\documentclass[12pt, centerh1]{article}
\usepackage{natbib}
\usepackage[margin=1in]{geometry}
\usepackage{amsmath,natbib,multirow}
\usepackage{amsfonts,mathrsfs,moreverb,lipsum}
\usepackage{graphicx,colonequals}
\usepackage{fixltx2e}
\usepackage{color}
\usepackage{caption}
\usepackage{subcaption}
\usepackage{mathtools}
\usepackage{pbox}
\usepackage{booktabs}
\usepackage{dsfont}
\usepackage{authblk}
\usepackage{url}
\newcommand{\tsub}{\textsubscript}
\newcommand\Tstrut{\rule{0pt}{2.6ex}}

\newcommand{\bfx}{{\bf x}}

\newcommand{\pig}{\pi_g}
\newcommand{\bvtheta}{{\boldsymbol \vartheta}}
\newcommand{\balp}{{\boldsymbol \alpha}}
\newcommand\numberthis{\addtocounter{equation}{1}\tag{\theequation}}

\author[1]{Michael P.B. Gallaugher\thanks{Corresponding author. Email: gallaump@mcmaster.ca}}
\author[1]{Paul D. McNicholas}

\affil[1]{Department of Mathematics and Statistics, McMaster University, Ontario, Canada}

\title{Clustering and Semi-Supervised Classification for Clickstream Data via Mixture Models}

\begin{document}

\maketitle

\begin{abstract}
Finite mixture models have been used for unsupervised learning for some time, and their use within the semi-supervised paradigm is becoming more commonplace. Clickstream data is one of the various emerging data types that demands particular attention because there is a notable paucity of statistical learning approaches currently available. A mixture of first-order continuous time Markov models is introduced for unsupervised and semi-supervised learning of clickstream data. This approach assumes continuous time, which distinguishes it from existing mixture model-based approaches; practically, this allows account to be taken of the amount of time each user spends on each webpage. The approach is evaluated, and compared to the discrete time approach, using simulated and real data. 
\end{abstract}

\section{Introduction}
Clickstream data present an important means of investigating users' internet behaviour. Unsupervised classification, a.k.a.\ clustering or cluster analysis, or semi-supervised classification of such data can be very useful in many different areas of endeavour. Examples can be found in areas as diverse as online marketing and anti-terrorism. Early examples of clustering clickstream data can be found in the work of \cite{banerjee00,banerjee01}, which looked at concept based clustering, and longest common sequences respectively. Other examples clustering and classification of clickstreams can be found in \cite{montgomery04}, \cite{aggarwal03} and \cite{wei12}; notably, none of these approaches draw on mixture models. 

The first use of mixture models for clustering for clickstreams can be found in \cite{cadez03}, who considered a mixture of first-order Markov models. One problem, as mentioned in \cite{cadez03}, is that the number of parameters can become very high when the number of website categories is very large. To alleviate this potential problem, \cite{Melny16} looked at bi-clustering of the clickstreams and the states to effectively reduce the number of states. Although this was shown to be successful in simulations, in the real data analyses, only two states were grouped together.

Herein, a mixture of first-order continuous time Markov models is introduced for unsupervised and semi-supervised classification of clickstream data. Specifically, the type of clickstream data considered herein would come from a website with multiple categories, such as {\tt amazon.com}, or a news website with categories such as weather, breaking news, sports, etc., with the clickstreams recording the movement of a user from one category to another, as well as the amount of time spent in each category. In practice, the incorporation of continuous time may be desirable in detecting the true underlying group structure for internet users. Consider the example of monitoring potentially inappropriate or criminal behaviour: our approach allows for the fact that an internet user might accidentally click on a link to, or be redirected to, inappropriate content and then immediately exit the site. When considering a discrete time Markov chain, the user would have been recorded as entering the inappropriate site, and could possibly be flagged as a problematic user. However, if the amount of time spent on the website can be taken into consideration, this may not be classified as suspicious activity. In the case of online shopping, a user could again click on the wrong category but immediately switch. Again, using a discrete time model would not be able to take this into account, and could lead to incorrect product suggestions.

The basis of the proposed methodology rests on the work done in \cite{Albert60}, who considered the estimation of the infinitesimal generator in a continuous time Markov model for a single component. Employing this in the mixture-model context using the EM algorithm is where the novelty lies.  
The balance of this paper is laid out as follows. In Section~\ref{background}, we look at a background of model-based clustering as well as the mixture of first-order Markov models and the extension to semi-supervised classification. In Section~\ref{method}, we discuss the extension to continuous time. In Section~\ref{analyses}, we look at three different simulations as well as dataset with simulated time points. Finally, we end with a discussion in Section~\ref{dis}.

\section{Background} \label{background}
\subsection{Mixture Models and Learning Therefrom}
A finite mixture model is a convex linear combination of component densities $f_1(\bfx~|~{\boldsymbol \theta}_1),\ldots,f_G(\bfx~|~{\boldsymbol \theta}_g)$. The density of a finite mixture model can be written
$$
f(\bfx~|~\bvtheta)=\sum_{g=1}^{G}\pig f_g(\bfx~|~{\boldsymbol \theta}_g),
$$
where $\bvtheta=\{\pi_1,\pi_2,\ldots,\pi_G,{\boldsymbol \theta}_1,{\boldsymbol \theta}_2,\ldots,{\boldsymbol \theta}_G\}$ and $\pig>0$, with $\sum_{g=1}^G\pi_g=1$, is the $g$th mixing proportion.
\cite{mcnicholas16a} traces the origin of the relationship between mixture models and unsupervised learning to the question of defining a cluster, where \cite{tiedeman55} and \cite{wolfe63} suggest using mixture models for this purpose. Some early work using (Gaussian) mixture models for clustering can be found in \cite{wolfe65}, \cite{baum70}, \cite{scott71} and \cite{orchard72}. The relative simplicity of the Gaussian mixture model, i.e., where each $f_g(\bfx~|~{\boldsymbol \theta}_g)$ is Gaussian, made it the dominant approach until relatively recently. In the last few years, however, there has been significant interest in the use of non-Gaussian mixture models for unsupervised and semi-supervised learning, e.g., work on mixtures of asymmetric distributions by \cite{lin10}, \cite{vrbik12,vrbik14}, \cite{lee14}, \cite{franczak14}, \cite{ohagan16}, \cite{smcnicholas17}, \cite{tang18}, and \cite{tortora19}. The, thus far limited, work on mixtures of Markov models for streaming data \citep{cadez03,Melny16} is another example of departure from Gaussian mixtures. 

One common application of mixture models is in the area of clustering and classification \citep[see][for a review]{mcnicholas16b}. In a typical classification problem, some observations have known group memberships, whereas some observations are unlabelled. Semi-supervised classification makes use of both the labelled and unlabelled points when building a classifier. In clustering, it is often the case that all observations are unlabelled; however, clustering can also be used when some observations are labelled. In this latter case, clustering may be performed in two different circumstances: when the labelled observations are treated as unlabelled or when the labelled observations are discarded. When clustering is performed using a mixture model, it is known as model-based clustering.

\subsection{Mixtures of First-Order Markov Models}
\cite{cadez03} and \cite{Melny16} consider a mixture of first-order Markov models to cluster clickstreams. Consider a website consisting of many different webpages that can be accessed from one of $J$ categories. The clickstream of interest is given by the transitions from one category to another. Suppose $N$ clickstreams are observed from a population with $G$ types. Now, assume that $N_1$ of these clickstreams have unknown labels and denote these clickstreams by ${\bf x}_i^{(1)}=(x_{i1}^{(1)},x_{i2}^{(1)},\ldots,x_{iL_i}^{(1)})', i\in\{1,2,\ldots,N_1\}$, and $N_2$ of these are labelled denoted similarly by ${\bf x}_i^{(2)}=(x_{i1}^{(2)},x_{i2}^{(2)},\ldots,x_{iL_i}^{(2)})', i\in\{1,2,\ldots,N_2\}$, where $N_2=N-N_1$ and $L_i$ is the length of clickstream $i$. For notational purposes, note that ${\bf x}_i^{(1)}$ is an $L_i$-dimensional vector of the states for the unlabelled clickstream $i$, and that each element can take values in the state space, which corresponds to the number of categories. For example, if there are 7 categories on a website, each element of ${\bf x}_i^{(1)}$ can take values in $\{1,2,\ldots,7\}$. The same applies to the labelled observations~${\bf x}_i^{(2)}$.

The one-step transition matrix for group $g$ is
$$
{\boldsymbol \Lambda}_g=\left(
\begin{array}{cccc}
\lambda_{g11}&\lambda_{g12}&\cdots& \lambda_{g1J}\\
\lambda_{g21}&\lambda_{g22}&\cdots& \lambda_{g2J}\\
\vdots & \vdots &\ddots &\vdots\\
\lambda_{gJ1}&\lambda_{gJ2}&\cdots & \lambda_{gJJ}
\end{array}
\right).
$$
Now, define the initial probabilities $\alpha_{gx_{i1}}=P(X_{i1}=x_{i1}|\bfx_i \mbox{ is in group } g),$ for $i=1,2,\ldots,N$. For ease of notation, and recalling that $x_{i1}\in\{1,2,\ldots,J\}$, denote an initial probability vector  for each group $g$ by $\balp_g=(\alpha_{g1},\alpha_{g2},\ldots,\alpha_{gJ})$.
Finally, for ease of notation, denote the total number of transitions from state $j$ to state $k$ for unlabelled clickstream $i$ by $n_{ijk}^{(1)}$ and likewise for labelled clickstream $i$ by $n_{ijk}^{(2)}$.

The observed likelihood is given by
\begin{equation}
\begin{split}
\mathcal{L}\tsub{\tiny obs}(\bvtheta|\mathcal{D}\tsub{\tiny \text{o}})&=
\underbrace{\prod_{i=1}^{N_1}\sum_{g=1}^G\left\{\pig\left[\prod_{j=1}^J\alpha_{gj}^{I(x_{i1}=j)}\right]\left[\prod_{j=1}^J\prod_{k=1}^J\lambda_{gjk}^{n_{ijk}}\right]\right\}}_{\text{Unlabelled Observations}} \\ &\qquad\qquad\times \underbrace{\prod_{i=1}^{N_2}\prod_{g=1}^G\left\{\pig\left[\prod_{j=1}^J\alpha_{gj}^{I(x_{i1}=j)}\right]\left[\prod_{j=1}^J\prod_{k=1}^J\lambda_{gjk}^{n_{ijk}}\right]\right\}^{z_{ig}^{(2)}}}_{\text{Labelled Observations}},
\end{split}
\end{equation}
where $\mathcal{D}\tsub{\tiny \text{O}}$ is the observed data and 
$$
z_{ig}^{(2)}=\left\{
\begin{array}{ll}
1 & \mbox{if labelled observation } i \mbox{ is in group } g,\\
0 & \mbox{otherwise.}
\label{eq:zig}
\end{array}
\right.
$$

The expectation-maximization (EM) algorithm \citep{dempster77} is used for parameter estimation. The EM algorithm works with the complete-data log-likelihood, i.e., the log-likelihood of the observed data together with the missing data. On the E-step, the expected value of the complete-data log-likelihood is computed and, on the M-step, it is maximized conditional on the current parameter estimates. The E- and M-steps are iterated until some stopping criterion is satisfied. Defining the latent $z_{ig}^{(1)}$ to be the group indicators for the unlabelled observations, analogous to $z_{ig}^{(2)}$ in \eqref{eq:zig}, the complete-data likelihood in this case can be written
\begin{equation*}
\mathcal{L}\tsub{\tiny \text{c}}=\prod_{m=1}^2\prod_{i=1}^{N_m}\prod_{g=1}^G\left\{\left[\prod_{j=1}^J\alpha_{gj}^{I(x_{i1}=j)}\right]\left[\prod_{j=1}^J\prod_{k=1}^J\lambda_{gjk}^{n_{ijk}}\right]\right\}^{z_{ig}^{(m)}}.
\end{equation*}
In this particular case, where the only latent variables in the EM are the $z_{ig}^{(1)}$ values, the EM algorithm can be outlined as follows.
\newline
\newline
{\bf Initialization}: Initialize the parameters $\pig$, $\balp_g$, and~${\boldsymbol \Lambda}_g$ for all $g=1,\ldots,G$. \newline \newline
{\bf E Step}: Update each $\hat{z}_{ig}^{(1)}$ by calculating
$$ 
\hat{z}_{ig}^{(1)}= \frac{\hat{\pi}_g\left[\prod_{j=1}^J\hat{\alpha}_{gj}^{I(x_{i1}^{(1)}=j)}\right]\left[\prod_{j=1}^J\prod_{k=1}^J\hat{\lambda}_{gjk}^{n_{ijk}^{(1)}}\right]}{\sum_{g=1}^G\hat{\pi}_g\left[\prod_{j=1}^J\hat{\alpha}_{gj}^{I(x_{i1}^{(1)}=j)}\right]\left[\prod_{j=1}^J\prod_{k=1}^J\hat{\lambda}_{gjk}^{n_{ijk}^{(1)}}\right]}.
$$
\newline 
{\bf M Step}: Update the parameter estimates via:
\begin{subequations}
\begin{equation}
\hat{\pi}_g=\frac{\sum_{m=1}^2\sum_{i=1}^{N_m} \hat{z}_{ig}^{(m)}}{N},
\label{eq:pigup}
\end{equation}
\begin{equation}
\hat{\alpha}_{gj}=\frac{\sum_{m=1}^2\sum_{i=1}^{N_m} \hat{z}_{ig}I(x_{i1}^{(m)}=j)}{\sum_{m=1}^2\sum_{i=1}^{N_m}\hat{z}_{ig}^{(m)}},
\label{eq:alphaup}
\end{equation}
\begin{equation}
\hat{\lambda}_{gjk}=\frac{\sum_{m=1}^{2}\sum_{i=1}^{N_{m}}\hat{z}_{ig}^{(m)}n^{(m)}_{ijk}}{\sum_{m=1}^2\sum_{i=1}^{N_m}\sum_{k'=1}^J\hat{z}_{ig}^{(m)}n^{(m)}_{ijk'}}.
\end{equation}
\end{subequations}
Note that unsupervised classification (clustering) falls out as the special case when $N_1=N$.

\section{Methodology}\label{method}
\subsection{Mixture of First-Order Continuous Time Markov Models}
We now discuss an extension of the methodology presented in \cite{cadez03} and \cite{Melny16} to take into account the amount of time spent in each category. Consider the same scenario as before, except this time we also observe a sequence of times spent in each state before transferring to another state; denote this by ${\bf t}_i^{(m)}=(t_{i1}^{(m)},t_{i2}^{(m)},\ldots,t_{iL_i}^{(m)})$ for $i=1,\ldots,N$, where $m=1$ corresponds to the unlabelled observations and $m=2$ corresponds to the labelled observations. It is important to note that, unlike the discrete time case, no transitions are made to the same state in continuous time.
These data can be modelled using a mixture of continuous time Markov chains, with infinitesimal generators
$$
{\bf Q}_g=\left(
\begin{array}{cccc}
q_{g11}&q_{g12}&\cdots& q_{g1J}\\
q_{g21}&q_{g22}&\cdots& q_{g2J}\\
\vdots & \vdots &\ddots &\vdots\\
q_{gJ1}&q_{gJ2}&\cdots & q_{gJJ}
\end{array}
\right),
$$
where
$
q_{gjk}\ge0 \mbox{ for } j\ne k$ and $  q_{gjj}=-\sum_{k\ne j}q_{gjk} \mbox{ for } g \in\{1,2,\ldots,G\}$.
The first item we note here is that the underlying transition probabilities are given by 
\begin{equation*}\begin{split}
P(X_{i(l+1)}^{(m)}=x_{i(l+1)}^{(m)}|X_{il}^{(m)}=x_{il}^{(m)}, &z_{ig}^{(m)}=1)=-\frac{q_{gx_{il}^{(m)}x_{i(l+1)}^{(m)}}}{q_{gx_{il}^{(m)}x_{il}^{(m)}}}.
\end{split}\end{equation*}
The second is that the $T_{il}^{(m)}$ are independent and
$$
T_{il}^{(m)}|(X_{il}^{(m)}=x_{il}^{(m)},z_{ig}^{(m)}=1)\sim \mbox{Exp}(-q_{gx_{il}^{(m)}x_{il}^{(m)}}),
$$ where $\mbox{Exp}(a)$ denotes an exponential distribution with rate~$a$. We denote the initial probability vector by $\balp_g$, as before.

\cite{Albert60} presents a detailed background for the theory of continuous time Markov chains, and also discusses the likelihood function for a sample of continuous time Markov chains for one component. Modifying this likelihood function for use in the mixture model context with multiple components, we obtain the likelihood
{\small\begin{align*}
&\mathcal{L}\tsub{\tiny obs}({\boldsymbol \vartheta}|\mathcal{D}\tsub{\tiny o})=\prod_{i=1}^{N_1}\sum_{g=1}^{G}\left\{\pi_g\left[\prod_{j=1}^{J}\alpha_{gj}^{I(x_{i1}^{(1)}=j)}\right]\left[\prod_{j=1}^J\prod_{k\ne j}^Jq_{gjk}^{n_{ijk}^{(1)}}\right]\left[-\prod_{j=1}^{J}q_{gjj}^{I(x_{iL_i}^{(1)}=j)}\right] \exp\left[\sum_{j=1}^{J}\sum_{l=1}^{L}q_{gjj}t_{il}^{(1)}I(x_{il}^{(1)}=j)\right]\right\}
 \\&\qquad\qquad \times \prod_{i=1}^{N_2}\prod_{g=1}^{G}\left\{\pi_g\left[\prod_{j=1}^{J}\alpha_{gj}^{I(x_{i1}^{(2)}=j)}\right]\left[\prod_{j=1}^J\prod_{k\ne j}^Jq_{gjk}^{n_{ijk}^{(2)}}\right]\left[-\prod_{j=1}^{J}q_{gjj}^{I(x_{iL_i}^{(2)}=j)}\right]\exp\left[\sum_{j=1}^{J}\sum_{l=1}^{L}q_{gjj}t_{il}^{(2)}I(x_{il}^{(2)}=j)\right] \right\}^{z_{ig}^{(2)}}
\end{align*}}
and the complete-data log-likelihood is
{\small \begin{align*}
\ell\tsub{\tiny c}&=\sum_{m=1}^2\sum_{i=1}^{N_m}\sum_{g=1}^{G}z_{ig}^{(m)}\left[\log\pig+\sum_{j=1}^J I(x_{i1}^{(m)}=j)\log\alpha_{gj}+\sum_{j=1}^J\sum_{k\ne j}^Jn_{ijk}^{(m)}\log q_{gjk}\right.\\ &\left.\qquad\qquad\qquad\qquad\qquad+\sum_{j=1}^J I(x_{iL_i}^{(m)}=j)\log(-q_{gjj})+\sum_{j=1}^J\sum_{l=1}^{L_i}q_{gjj}t_{il}^{(m)}I(x_{il}^{(m)})\right].
\numberthis \label{eq:cll}
\end{align*}}

In the E step, we update the latent indicator variables $z_{ig}^{(1)}$, and these are the only latent variables for this EM algorithm. At iteration $s+1$, this update is given by

\begin{equation}
\hat{z}_{ig}^{(1)}=\frac{h(\hat{\pi}_g,\hat{\balp}_g,\hat{{\bf Q}}_g,\bfx_i^{(1)},{\bf t}_i^{(1)})}{\sum_{g'=1}^Gh(\hat{\pi}_{g'},\hat{\balp}_{g'},\hat{{\bf Q}}_{g'},\bfx_i^{(1)},{\bf t}_i^{(1)})},
\label{eq:zup}
\end{equation}
where
\begin{align*}
h(\hat{\pi}_g,\hat{\balp}_g,\hat{{\bf Q}}_g,\bfx_i^{(1)},{\bf t}_i^{(1)}) &= \hat{\pi}_g\left[\prod_{j=1}^{J}(\hat{\alpha}_{gj})^{I(x_{i1}^{(1)}=j)}\right]\left[\prod_{j=1}^{J}(-\hat{q}_{gjj})^{I(x_{iL}^{(1)}=j)}\right] \left[\prod_{j=1}^J\prod_{k\ne j}^J(\hat{q}_{gjk})^{n_{ijk}^{(1)}}\right] \\ 
&\quad\times \exp\left\{\sum_{j=1}^{J}\sum_{l=1}^{L_i}\hat{q}_{gjj}t_{il}^{(1)}I(x_{il}^{(1)}=j)\right\}.
\end{align*}
In the M step, we update our parameters, $\hat{\pi}_g$, $\hat{\balp}_g$ and $\hat{{\bf Q}}_g$. The updates for the $\hat{\pi}_g$ and $\hat{\balp}_g$ are the same as those in the discrete case, see \eqref{eq:pigup} and \eqref{eq:alphaup}.
We also update each $\hat{{\bf Q}}_g$, and these updates are given by
$$
\hat{q}_{gjk}=
\left\{
\begin{array}{cc}
\frac{\sum_{i=1}^Nz_{ig}^{(m)}n_{ijk}}{\hat{\lambda}_{gj}} & \mbox{if } j\ne k,\\
-\sum_{k\ne j}\hat{q}_{gjk} & \mbox{if } k=j,
\end{array}
\right.
$$
where $\hat{\lambda}_{gj}=a/b$ with
\begin{equation*}\begin{split}
a&=\sum_{m=1}^2\sum_{i=1}^{N_m}\left[\sum_{l=1}^{L_i}\hat{z}_{ig}^{(m)}t_{il}^{(m)}I(x_{il}^{(m)}=j) +\sum_{k\ne j}^J\hat{z}_{ig}^{(m)}n_{ijk}^{(m)}\right],\\
b&=\sum_{m=1}^2\sum_{i=1}^{N_m}\hat{z}_{ig}^{(m)}I(x_{iL_i}^{(m)}=j)+\sum_{m=1}^2\sum_{i=1}^{N_m}\sum_{k\ne j}^J \hat{z}_{ig}^{(m)}n_{ijk}^{(m)}.
\end{split}\end{equation*}
Notably, the number of free parameters in this continuous time model are the same as those in the discrete time model.

\subsection{Initialization, Convergence, and Computational Issues}
We used the emEM algorithm described in \cite{biernacki03} and used by \cite{Melny16}. This procedure consists of a short em step in which the EM algorithm is performed for a small number of iterations with different random starting values. The initialization which attains the greatest likelihood after the short em algorithm is then used for the full EM algorithm. In our analyses, we used 50 random starting values and ran the short em algorithm for five iterations. 

We use a criterion based on the Aitken acceleration \citep{aitken26} to determine convergence. 
The Aitken acceleration at iteration $k$ is
\begin{equation}\label{eqn:aa}
a^{(k)} = \frac{l^{(k+1)}-l^{(k)}}{l^{(k)}-l^{(k-1)}},
\end{equation}
where $l^{(k)}$ is the observed log-likelihood at iteration $k$. The quantity in \eqref{eqn:aa} can be used to derive an asymptotic estimate of the log-likelihood at iteration $k+1$: $$l_{\infty}^{(k+1)} = l^{(k)} + \frac{1}{1-a^{(k)}}(l^{(k+1)}-l^{(k)}),$$ and the EM algorithm can stopped when $l_{\infty}^{(k+1)}-l^{(k)} < \epsilon$ \citep[see][]{bohning94, lindsay95,mcnicholas10a}.

We note that calculating the updates for $\hat{z}^{(1)}_{ig}$ using \eqref{eq:zup} can lead to computational problems. This is due to the calculation of $h(\cdot)$ being computationally equal to 0 for all groups $g$, which occurs when $\log h(\cdot)$ becomes large and negative, followed by exponentiating this large negative value to get the value of $h(\cdot)$. Taking this into account, we can rewrite this update as
\begin{align*}
\frac{1}{\hat{z}^{(1)}_{ig}}&=\sum_{g'=1}^{G}\exp\left\{\log\left(\frac{\hat\pi_{g'}}{\hat\pi_g}\right)+\sum_{j=1}^{J}I(x_{i1}=j)\log\left(\frac{\hat\alpha_{g'j}}{\hat\alpha_{gj}}\right)+\sum_{j=1}^{J}\sum_{k=1}^{J}n_{ijk}\log\left(\frac{\hat{q}_{g'jk}}{\hat{q}_{gjk}}\right)\right.\\ &\qquad\qquad\qquad+\left. \sum_{j=1}^{J}I(x_{iL_i})\log\left(\frac{\hat{q}_{g'jj}}{\hat{q}_{gjj}}\right)+\sum_{j=1}^{J}\sum_{l=1}^{L_{i}}I(x_{il}=j)t_{il}(\hat{q}_{g'jj}-\hat{q}_{gjj})\right\},
\end{align*}
which allows the $\hat{z}^{(1)}_{ig}$ to be computed directly instead of having to calculate the $h(\cdot)$ functions for each group separately.

A second computational issue, as discussed in \cite{Melny16}, is the case where there are no transitions present in the data between two states, i.e., $n_{ijk}=0$ for all $i$. In this case, the estimates for $q_{gjk}$ would be zero for all $g$. Firstly, this is a problem because we can make the reasonable assumption that all states communicate with each other, making an estimate of zero unrealistic. Secondly, this would cause problems with the calculation of the likelihood. We, therefore, set a lower bound of $10^{-6}$ for all parameter values.

\subsection{Selecting the Number of Groups}
For the clustering case, in general, we do not know the number of groups {\it a priori} and, hence, we need to choose an appropriate number of groups. There have been a few different selection criteria proposed in the literature; however, the Bayesian information criterion \citep[BIC;][]{schwarz78} remains the most popular approach. Moreover, the BIC was used by \cite{Melny16} for the discrete time model. The BIC can be written 
$$
\mbox{BIC}=2\ell_{\tiny \text{obs}}(\bvtheta|D)-p\log N,
$$
where $p$ is the number of free parameters.

An alternative to the BIC, that is also quite well-known in the literature, is the integrated complete likelihood \citep[ICL;][]{biernacki00}. The ICL can be approximated via
$$
\mbox{ICL}\approx \mbox{BIC}+2\sum_{i=1}^{N_1}\sum_{g=1}^G\mbox{MAP}(\hat{z}_{ig}^{(1)})\log(\hat{z}_{ig}^{(1)}),
$$
where 
$$
\text{MAP}(\hat{z}_{ig}^{(1)})=\left\{
\begin{array}{lll}
1 && \text{if arg max}_{h=1,\ldots,G}\{\hat{z}_{ig}^{(1)}\}=g,\\
0 && \text{otherwise}.
\end{array}
\right.
$$
The second term in the formula for the ICL is an entropy term and so the ICL can be considered as a penalized BIC, where the penalty is based on the uncertainty in the estimated classifications. We note here that only the results when using the BIC are shown in the simulations because the results using the ICL are almost identical.

In the case of semi-supervised classification, it is possible that one or more subgroups are not represented in the known labels, in which case the true number of groups could be set to $H>G$. It is also possible that the known labels actually represent an incorrect (subgroup) structure, in which case the true number of groups could be $H<G$. Despite these two possibilities, we will 
assume that $H=G$ in our analyses.

\section{Analyses}\label{analyses}
\subsection{Simulation 1}
In the first simulation, we simulated from two groups with infinitesimal generators
$$
{\bf Q}_1=\left(
\begin{array}{rrrrr}
-0.100 & 0.050 & 0.020 & 0.020 & 0.010 \\ 
  0.100 & -1.000 & 0.200 & 0.100 & 0.600 \\ 
  0.020 & 0.050 & -0.100 & 0.005 & 0.025 \\ 
  0.050 & 0.050 & 0.050 & -1.000 & 0.850 \\ 
  0.006 & 0.004 & 0.050 & 0.040 & -0.100 \\ 
\end{array}\right),
$$
$$
{\bf Q}_2=\left(
\begin{array}{rrrrr}
-0.100 & 0.001 & 0.009 & 0.015 & 0.075 \\ 
  0.700 & -1.000 & 0.200 & 0.050 & 0.050 \\ 
  0.010 & 0.005 & -0.100 & 0.030 & 0.055 \\ 
  0.400 & 0.400 & 0.100 & -1.000 & 0.100 \\ 
  0.030 & 0.030 & 0.020 & 0.020 & -0.100 \\ 
\end{array}
\right).
$$
We took sample sizes of $N\in\{50,100,200,400\}$ with clickstream lengths $L$ ranging from 4 to 25 and 25 to 100. We also considered two cases with equal proportions, $\pi_1=\pi_2=0.5$ (Simulation 1A) and $\pi_1=0.2, \pi_2=0.8$ (Simulation 1B). The purpose of the second case is that, when looking for suspicious behaviour, it is highly likely that there are fewer suspicious users than regular users.

In these simulations, we have large separation in the underlying transition probabilities; however, because the diagonal elements are identical between groups, there is no separation in the average amount of time spent in each category. The results for Simulations 1A and B for both the discrete and continuous time models are summarized in Tables~\ref{tab:Sim1a} and \ref{tab:Sim1b}. After fitting the model for $G=1,2,\ldots,5$, we consider the number of times each $G$ was chosen using the BIC as well as the average adjusted Rand index \citep[ARI;][]{hubert85} and the associated standard deviation. The ARI compares two different partitions, in our case comparing the true labels and the estimated labels, and has a value of 1 if there is perfect agreement. The expected value of the ARI under random class agreement is 0.

In this case, we see that the results for the continuous and discrete models are almost identical, with only slight variations in the ARI between the two methods. The BIC in all cases with low values of $L$, correctly finds two groups for both the continuous and discrete models. When $L$ is increased, there is a very slight chance of overfitting the number of groups. It is interesting to note that a higher sample size does not affect the classification performance as much as a longer length of the clickstream. Finally, very little difference is seen when changing the mixing proportions. The similar results between the discrete and continuous time models illustrate the ability of the continuous time model to effectively detect group structure based solely on differences in transition probabilities.
\begin{table}[!ht]
\centering
\caption{Summary of the results from Simulation 1A  ($\pi_1=\pi_2=0.5$).}
\begin{tabular}{llrrrrrc}
  \hline
\multicolumn{8}{c}{$L$ from 4 to 25}\\
\hline
 Sample Size&Model & $G$=1 & $G$=2 & $G$=3 & $G$=4 & $G$=5 & $\overline{\text{ARI}}$  (sd) \Tstrut \\
\hline
\multirow{2}{*}{$N$=50}&Continuous & 0 & 100 & 0 & 0 & 0 & 0.935 (0.069)\\
 &Discrete & 0 & 100 & 0 & 0 & 0 & 0.942 (0.06) \\   \hline 
\multirow{2}{*}{$N$=100} &  Continuous & 0 & 100 & 0 & 0 & 0 & 0.955(0.045) \\ 
 & Discrete & 0 & 100 & 0 & 0 & 0 & 0.955(0.043) \\     \hline
\multirow{2}{*}{$N$=200}  &  Continuous & 0 & 100 & 0 & 0 & 0 & 0.958(0.029) \\ 
 & Discrete & 0 & 100 & 0 & 0 & 0 & 0.958(0.029) \\ \hline
\multirow{2}{*}{$N$=400}   &  Continuous & 0 & 100 & 0 & 0 & 0 & 0.950(0.026) \\ 
 & Discrete & 0 & 100 & 0 & 0 & 0 & 0.950(0.025) \\
  \hline
  \multicolumn{8}{c}{$L$ from 25 to 100}\\
  \hline
   Sample Size&Model & $G$=1 & $G$=2 & $G$=3 & $G$=4 & $G$=5 & $\overline{\text{ARI}}$  (sd) \Tstrut \\
\hline
\multirow{2}{*}{$N$=50}  &Continuous & 0 & 98 & 2 & 0 & 0 & 0.995 (0.032)\\
 &Discrete & 0 & 98 & 2 & 0 & 0 & 0.995 (0.035) \\   \hline 
\multirow{2}{*}{$N$=100} &  Continuous & 0 & 99 & 1 & 0 & 0 & 0.999(0.012) \\ 
 & Discrete & 0 & 96 & 4 & 0 & 0 & 0.994(0.036) \\     \hline
\multirow{2}{*}{$N$=200}  &  Continuous & 0 & 98 & 2 & 0 & 0 & 0.999(0.008) \\ 
 & Discrete & 0 & 98 & 2 & 0 & 0 & 0.997(0.022) \\ \hline
\multirow{2}{*}{$N$=400}   &  Continuous & 0 & 99 & 1 & 0 & 0 & 0.999(0.004) \\ 
 & Discrete & 0 & 95 & 5 & 0 & 0 & 0.998(0.012) \\
  \hline
\end{tabular}
\label{tab:Sim1a}
\end{table}
\begin{table}[!ht]
\centering
\caption{Summary of the results from Simulation 1B  ($\pi_1=0.2,\pi_2=0.8$).}
\begin{tabular}{llrrrrrc}
\hline
\multicolumn{8}{c}{$L$ from 4 to 25}\\
\hline
 Sample Size&Model & $G$=1 & $G$=2 & $G$=3 & $G$=4 & $G$=5 & $\overline{\text{ARI}}$  (sd) \Tstrut \\
\hline
\multirow{2}{*}{$N$=50}&Continuous & 0 & 100 & 0 & 0 & 0 & 0.940 (0.070)\\
 &Discrete & 0 & 100 & 0 & 0 & 0 & 0.948 (0.067) \\   \hline 
\multirow{2}{*}{$N$=100} &  Continuous & 0 & 100 & 0 & 0 & 0 & 0.955(0.042) \\ 
 & Discrete & 0 & 100 & 0 & 0 & 0 & 0.957(0.041) \\     \hline
\multirow{2}{*}{$N$=200}  &  Continuous & 0 & 100 & 0 & 0 & 0 & 0.960(0.032) \\ 
 & Discrete & 0 & 100 & 0 & 0 & 0 & 0.957(0.032) \\ \hline
\multirow{2}{*}{$N$=400}   &  Continuous & 0 & 100 & 0 & 0 & 0 & 0.957(0.024) \\ 
 & Discrete & 0 & 100 & 0 & 0 & 0 & 0.958(0.025) \\
  \hline
  \multicolumn{8}{c}{$L$ from 25 to 100}\\
  \hline
   Sample Size&Model & $G$=1 & $G$=2 & $G$=3 & $G$=4 & $G$=5 & $\overline{\text{ARI}}$  (sd) \Tstrut \\
\hline
 \multirow{2}{*}{$N$=50} &Continuous & 0 & 91 & 9 & 0 & 0 & 0.960 (0.14)\\
 &Discrete & 0 & 91 & 9 & 0 & 0 & 0.960 (0.14) \\   \hline 
\multirow{2}{*}{$N$=100} &  Continuous & 0 & 99 & 1 & 0 & 0 & 0.999(0.004) \\ 
 & Discrete & 0 & 89 & 11 & 0 & 0 & 0.965(0.13) \\     \hline
\multirow{2}{*}{$N$=200}  &  Continuous & 0 & 96 & 4 & 0 & 0 & 0.992(0.056) \\ 
 & Discrete & 0 & 91 & 9 & 0 & 0 & 0.970(0.11) \\ \hline
\multirow{2}{*}{$N$=400}  &  Continuous & 0 & 96 & 4 & 0 & 0 & 0.996(0.026) \\ 
 & Discrete & 0 & 91 & 9 & 0 & 0 & 0.987(0.067) \\
  \hline
\end{tabular}
\label{tab:Sim1b}
\end{table}

\subsection{Simulation 2}
In this simulation, we once again look at clustering. This time, data are simulated from three different groups, with mixing proportions $\pi_1=\pi_2=\pi_3=1/3$ (Simulation 2A) and $\pi_1=0.2$, $\pi_2=0.4, \pi_3=0.4$ (Simulation 2B). There are seven states, $\balp_1$ is taken to be uniform, $\balp_2$ gives probability 0.1 for all states except state 7, which has probability of 0.4, and $\balp_3$ gives probability 0.1 to all states except state 3, which has probability 0.4. The infinitesimal generators are taken to be 
$$
{\bf Q}_1=\left(
\begin{array}{rrrrrrr}
-0.14 & 0.05 & 0.02 & 0.02 & 0.01 & 0.02 & 0.02 \\ 
  0.10 & -1.40 & 0.20 & 0.10 & 0.60 & 0.20 & 0.20 \\ 
  0.02 & 0.05 & -0.14 & 0.01 & 0.03 & 0.02 & 0.02 \\ 
  0.05 & 0.05 & 0.05 & -1.40 & 0.80 & 0.25 & 0.20 \\ 
  0.01 & 0.00 & 0.05 & 0.04 & -0.14 & 0.04 & 0.01 \\ 
  0.70 & 0.10 & 0.10 & 0.10 & 0.10 & -1.40 & 0.30 \\ 
  0.50 & 0.50 & 0.05 & 0.05 & 0.10 & 0.20 & -1.40 \\ 
  \end{array}
\right),
$$
$$
{\bf Q}_2=\left(
\begin{array}{rrrrrrr}
-1.40 & 0.40 & 0.30 & 0.15 & 0.15 & 0.25 & 0.15 \\ 
  0.02 & -0.14 & 0.03 & 0.02 & 0.03 & 0.03 & 0.01 \\ 
  0.30 & 0.50 & -1.40 & 0.10 & 0.10 & 0.20 & 0.20 \\ 
  0.01 & 0.01 & 0.01 & -0.14 & 0.05 & 0.03 & 0.03 \\ 
  0.01 & 0.01 & 0.04 & 0.05 & -0.14 & 0.02 & 0.02 \\ 
  0.70 & 0.05 & 0.15 & 0.05 & 0.15 & -1.40 & 0.30 \\ 
  0.05 & 0.05 & 0.01 & 0.01 & 0.01 & 0.01 & -0.14 \\ 
\end{array}
\right),
$$
$$
{\bf Q}_3=\left(
\begin{array}{rrrrrrr}
-1.40 & 0.20 & 0.70 & 0.20 & 0.10 & 0.10 & 0.10 \\ 
  0.60 & -1.40 & 0.20 & 0.20 & 0.20 & 0.10 & 0.10 \\ 
  0.10 & 0.10 & -1.40 & 0.80 & 0.10 & 0.10 & 0.20 \\ 
  0.05 & 0.03 & 0.03 & -0.14 & 0.01 & 0.01 & 0.01 \\ 
  0.05 & 0.05 & 0.01 & 0.01 & -0.14 & 0.01 & 0.02 \\ 
  1.00 & 0.02 & 0.03 & 0.02 & 0.03 & -1.40 & 0.30 \\ 
  0.20 & 0.20 & 0.20 & 0.20 & 0.20 & 0.40 & -1.40 \\ 
\end{array}
\right).
$$

In this case, there are two groups with similar underlying transition probabilities, but different amounts of time on average being spent in each state. The third group has a large amount of separation in the underlying transition probabilities in comparison to the first two. Also, the third group is defined by more time spent in states 4 and 5 on average than the rest of the states. From the results (Tables~\ref{tab:Sim2a} and \ref{tab:Sim2b}), we see that the continuous time model outperforms the discrete time model in all cases. Specifically, for short clickstream lengths, the BIC under-fits the true number of groups in all cases for the discrete model. Increasing $N$ for shorter clickstreams helps the discrete model a little, i.e., for small $N$, the discrete time model finds only one group but, for larger $N$, two groups are selected, which is closer to the true number of groups. Increasing the length of the clickstream also helps with selecting the correct number of groups for the discrete time model, but still requires a sample size of 600 to choose the correct number of groups in all cases. The continuous time model performs well for all values of $L$ and $N$---again, increasing the sample size is not as impactful as increasing the clickstream length. 
It is not surprising that the continuous time model outperforms the discrete time model in this case because there is very little separation in the underlying transition probabilities for two of the groups but the time spent in each state is fairly well separated between the three groups. 
Accordingly, the discrete model, being unable to take into account the amount of time in each state, is unable to distinguish between groups~1 and 2.
\begin{table}[!ht]
\centering
\caption{Summary of results for Simulation 2A ($\pi_1=\pi_2=\pi_3=1/3$).}
\begin{tabular}{llrrrrrc}
\hline
\multicolumn{8}{c}{$L$ from 4 to 25}\\
\hline
 Sample Size&Model & $G$=1 & $G$=2 & $G$=3 & $G$=4 & $G$=5 & $\overline{\text{ARI}}$  (sd) \Tstrut \\
\hline
\multirow{2}{*}{$N$=75}&Continuous & 0 & 0 & 100 & 0 & 0 & 0.934 (0.045)\\
 &Discrete & 34 & 66 & 0 & 0 & 0 & 0.302 (0.22) \\   \hline 
\multirow{2}{*}{$N$=150} &  Continuous & 0 & 0 & 100 & 0 & 0 & 0.952(0.026) \\ 
 & Discrete & 0 & 100 & 0 & 0 & 0 & 0.467(0.043) \\     \hline
\multirow{2}{*}{$N$=300}  &  Continuous & 0 & 0 & 100 & 0 & 0 & 0.958(0.021) \\ 
 & Discrete & 0 & 100 & 0 & 0 & 0 & 0.477(0.032) \\ \hline
\multirow{2}{*}{$N$=600}   &  Continuous & 0 & 0 & 100 & 0 & 0 & 0.956(0.014) \\ 
 & Discrete & 0 & 100 & 0 & 0 & 0 & 0.476(0.021) \\
  \hline
  \multicolumn{8}{c}{$L$ from 25 to 100}\\
  \hline
   Sample Size&Model & $G$=1 & $G$=2 & $G$=3 & $G$=4 & $G$=5 & $\overline{\text{ARI}}$  (sd) \Tstrut \\
\hline
\multirow{2}{*}{$N$=75}  &Continuous & 0 & 0 & 95 & 5 & 0 & 0.994 (0.026)\\
 &Discrete & 0 & 100 & 0 & 0 & 0 & 0.564 (0.004) \\   \hline 
\multirow{2}{*}{$N$=150} &  Continuous & 0 & 0 & 94 & 6 & 0 & 0.994(0.023) \\ 
 & Discrete & 0 & 89 & 11 & 0 & 0 & 0.603(0.10) \\     \hline
\multirow{2}{*}{$N$=300}  &  Continuous & 0 & 0 & 98 & 2 & 0 & 0.992(0.056) \\ 
 & Discrete & 0 & 2 & 98 & 0 & 0 & 0.872(0.051) \\ \hline
\multirow{2}{*}{$N$=600}   &  Continuous & 0 & 0 & 97 & 3 & 0 & 0.999(0.006) \\ 
 & Discrete & 0 & 0 & 100 & 0 & 0 & 0.882(0.020) \\
  \hline
\end{tabular}
\label{tab:Sim2a}
\end{table}
\begin{table}[!ht]
\centering
\caption{Summary of results for Simulation 2B ($\pi_1=0.2,\pi_2=0.4,\pi_3=0.4$).}
\begin{tabular}{llrrrrrc}
\hline
\multicolumn{8}{c}{$L$ from 4 to 25}\\
\hline
 Sample Size&Model & $G$=1 & $G$=2 & $G$=3 & $G$=4 & $G$=5 & $\overline{\text{ARI}}$  (sd) \Tstrut \\
\hline
\multirow{2}{*}{$N$=75}&Continuous & 0 & 0 & 100 & 0 & 0 & 0.935 (0.045)\\
 &Discrete & 30 & 70 & 0 & 0 & 0 & 0.382 (0.26) \\   \hline 
\multirow{2}{*}{$N$=150} &  Continuous & 0 & 0 & 100 & 0 & 0 & 0.948(0.030) \\ 
 & Discrete & 0 & 100 & 0 & 0 & 0 & 0.559(0.050) \\     \hline
\multirow{2}{*}{$N$=300}  &  Continuous & 0 & 0 & 100 & 0 & 0 & 0.953(0.023) \\ 
 & Discrete & 0 & 100 & 0 & 0 & 0 & 0.566(0.034) \\ \hline
\multirow{2}{*}{$N$=600}  &  Continuous & 0 & 0 & 100 & 0 & 0 & 0.953(0.015) \\ 
 & Discrete & 0 & 100 & 0 & 0 & 0 & 0.569(0.025) \\
  \hline
  \multicolumn{8}{c}{$L$ from 25 to 100}\\
  \hline
   Sample Size&Model & $G$=1 & $G$=2 & $G$=3 & $G$=4 & $G$=5 & $\overline{\text{ARI}}$  (sd) \Tstrut \\
\hline
\multirow{2}{*}{$N$=75}  &Continuous & 0 & 0 & 98 & 2 & 0 & 0.999 (0.011)\\
 &Discrete & 0 & 100 & 0 & 0 & 0 & 0.677 (0.007) \\   \hline 
\multirow{2}{*}{$N$=150} &  Continuous & 0 & 0 & 96 & 4 & 0 & 0.996(0.021) \\ 
 & Discrete & 0 & 99 & 1 & 0 & 0 & 0.682(0.031) \\     \hline
\multirow{2}{*}{$N$=300}  &  Continuous & 0 & 0 & 96 & 4 & 0 & 0.997(0.018) \\ 
 & Discrete & 0 & 29 & 71 & 0 & 0 & 0.842(0.10) \\ \hline
\multirow{2}{*}{$N$=600}   &  Continuous & 0 & 0 & 97 & 3 & 0 & 0.999(0.004) \\ 
 & Discrete & 0 & 0 & 100 & 0 & 0 & 0.915(0.018) \\
  \hline
\end{tabular}
\label{tab:Sim2b}
\end{table}

\subsection{Simulation 3}
In this simulation, semi-supervised classification was considered by simulating under the same circumstances as Simulation 1A (Simulation 3A) and Simulation 2A (Simulation 3B). Supervision levels of 20, 40 and 80\% were considered, and the average ARI values for observations considered unlabelled with standard deviations are shown in Tables \ref{tab:Sim3a} and \ref{tab:Sim3b}. 

For Simulation 3A, we again see that the results are very similar between the discrete and continuous time models. Increasing the level of supervision, unsurprisingly improves the classification performance for short clickstream lengths. When increasing the lengths of the clickstreams, perfect classification performance is achieved.
For Simulation 3B, again the clickstream model outperforms the discrete model. Moreover, increasing the clickstream length again improves the performance for both models, giving perfect classification for the continuous time model for all levels of supervision, and a much improved ARI for the discrete time model.
\begin{table}[!ht]
\centering
\caption{Average ARI values for unlabelled observations over 100 datasets, with standard deviations in parentheses, for Simulation 3A.}
\begin{tabular}{llccc}
  \hline
  \multicolumn{5}{c}{$L$ from 4 to 25}\\
\hline
Sample Size&Model & 20\% Supervision & 40\% Supervision& 80\% Supervision\Tstrut \\ 
\hline
\multirow{2}{*}{$N$=50}&Continuous&0.946(0.069)& 0.967(0.068)&0.984(0.079)\\
 &Discrete & 0.945(0.071)& 0.967(0.068) &0.984(0.079)\\   \hline 
 \multirow{2}{*}{$N$=100}&  Continuous&0.948(0.046)& 0.960(0.047)& 0.990(0.044) \\ 
 & Discrete&0.947(0.046)& 0.964(0.042)& 0.992(0.039)  \\     \hline
\multirow{2}{*}{$N$=200}  &  Continuous& 0.954(0.035)& 0.969(0.034)& 0.987(0.034)  \\ 
 & Discrete &0.954(0.034)&0.969(0.032)& 0.987(0.034) \\ \hline
\multirow{2}{*}{$N$=400}   &  Continuous & 0.963(0.020)& 0.971(0.022)& 0.991(0.021) \\ 
 & Discrete & 0.963(0.020)& 0.972(0.021)& 0.991(0.021)\\
  \hline
  \multicolumn{5}{c}{$L$ from 25 to 100}\\
  \hline
\multirow{2}{*}{$N$=50}  &Continuous &1.00(0.00)&1.00(0.00) &1.00(0.00)\\
 &Discrete & 1.00(0.00)&1.00(0.00) &1.00(0.00)\\   \hline 
\multirow{2}{*}{$N$=100} &  Continuous & 1.00(0.00)&1.00(0.00) &1.00(0.00) \\ 
 & Discrete &1.00(0.00)&1.00(0.00) &1.00(0.00)  \\     \hline
\multirow{2}{*}{$N$=200}  &  Continuous &1.00(0.00)&1.00(0.00) &1.00(0.00)   \\ 
 & Discrete &1.00(0.00)&1.00(0.00) &1.00(0.00)   \\ \hline
\multirow{2}{*}{$N$=400}   &  Continuous &1.00(0.00)&1.00(0.00) &1.00(0.00)  \\ 
 & Discrete & 1.00(0.00)&1.00(0.00) &1.00(0.00) \\
\hline
\end{tabular}
\label{tab:Sim3a}
\end{table}
\begin{table}[!ht]
\centering
\caption{Average ARI values for unlabelled observations over 100 datasets, with standard deviations in parentheses, for Simulation 3B.}
\begin{tabular}{llccc}
  \hline
  \multicolumn{5}{c}{$L$ from 4 to 25}\\
\hline
Sample Size&Model & 20\% Supervision & 40\% Supervision& 80\% Supervision\Tstrut \\ 
\hline
\multirow{2}{*}{$N$=75}&Continuous&0.959(0.042)& 0.968(0.045)& 0.988(0.050)\\
 &Discrete & 0.501(0.087)& 0.594(0.109)& 0.828(0.168)\\   \hline 
\multirow{2}{*}{$N$=150} &  Continuous&0.960(0.028)& 0.969(0.029)& 0.988(0.033) \\ 
 & Discrete&0.526(0.074)& 0.638(0.065)& 0.882(0.103)\\     \hline
\multirow{2}{*}{$N$=300}  &  Continuous&0.966(0.021)& 0.976(0.020)& 0.989(0.022)  \\ 
 & Discrete &0.578(0.053)& 0.670(0.058)& 0.890(0.072) \\ \hline
\multirow{2}{*}{$N$=600}   &  Continuous & 0.967(0.013)& 0.975(0.013)& 0.993(0.013) \\ 
 & Discrete &0.616(0.034)& 0.702(0.037)& 0.890(0.056)\\
  \hline
  \multicolumn{5}{c}{$L$ from 25 to 100}\\
  \hline
\multirow{2}{*}{$N$=75}  &Continuous &1.00(0.00)&1.00(0.00) &1.00(0.00)\\
 &Discrete &0.848(0.080)&0.899(0.069)& 0.977(0.071) \\   \hline 
\multirow{2}{*}{$N$=150} &  Continuous & 1.00(0.00)&1.00(0.00) &1.00(0.00) \\ 
 & Discrete &0.889(0.043)& 0.904(0.052)&0.973(0.051)  \\     \hline
\multirow{2}{*}{$N$=300}  &  Continuous &1.00(0.00)&1.00(0.00) &1.00(0.00)   \\ 
 & Discrete &0.906(0.033)& 0.927(0.038)& 0.978(0.033)  \\ \hline
\multirow{2}{*}{$N$=600} &  Continuous &1.00(0.00)&1.00(0.00) &1.00(0.00)  \\ 
 & Discrete &0.909(0.021)&0.928(0.020)& 0.972(0.027) \\
  \hline
\end{tabular}
\label{tab:Sim3b}
\end{table}

\section{Real Data Analysis}
The YouChoose clicks dataset \citep{youchoose} is now considered. This dataset considers a retail website with 12 states, labelled 1 to 12 along with a special category notated as ``S", which we relabelled to State 13. Missing values were recorded as 0, and observations with missing values in the clickstream are removed from the analysis. In addition, occasionally users went to a page for a particular brand labelled with an eight to ten digit number. Most clickstreams did not include a transition to one of these states, and for the purposes of our analysis, users that did are removed.  A time stamp is only given for when a user entered a state, and not when the user left the state; therefore, we had to remove the final state from the clickstream. Finally, we only considered clickstreams with at least two transitions, and time was measured in seconds. After cleaning, there are a total of 14561 clickstreams. We fit all models for one to twelve groups, and choose the best model based on the BIC. The BIC chooses three groups for both the discrete and continuous time models.

\begin{figure}[ht]
\centering
\includegraphics[width=\textwidth]{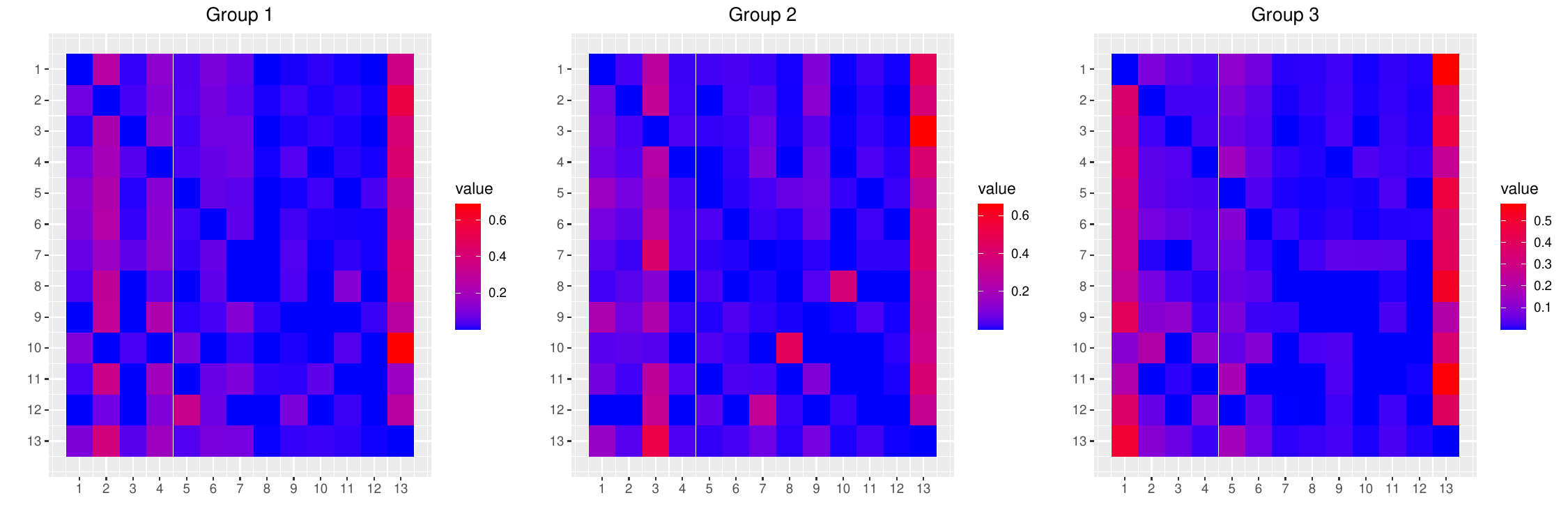}
\caption{Heat maps of estimated transition matrices for groups 1, 2, and 3 for the discrete time model.}
\label{fig:Dis}
\end{figure}

 In Figure \ref{fig:Dis}, heat maps of the estimated transition matrices for the three groups for the discrete time model are shown. In all three groups, users appear to have a high probability of transitioning to State 13. Considering that State 13 relates to a special or promotion, this might be expected behaviour. Group 3 appears to be categorized by users with a fairly strong tendency to transition to State 1 as well as State 13. In Group 2, there is fairly high probability of transitioning to State 3 from all states. Furthermore, there appears to be high transition probabilities between States 8 and 10 and vice versa.

\begin{figure}[!htb]
\centering
\includegraphics[width=\textwidth]{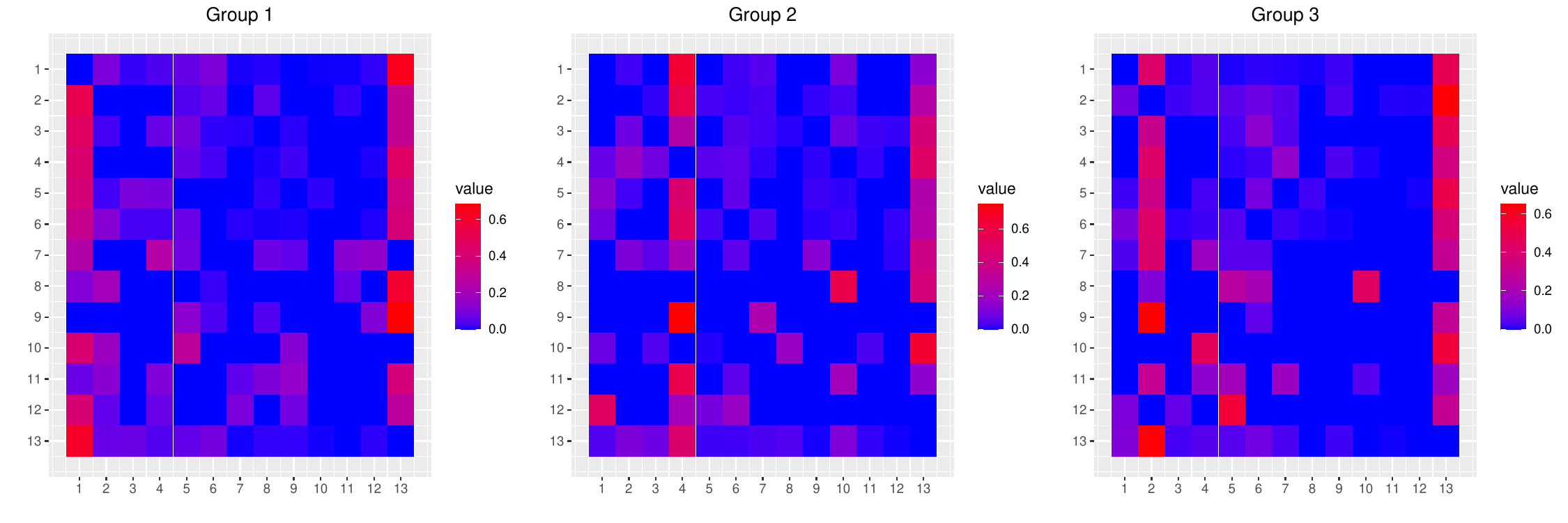}
\caption{Heat maps of estimated transition matrices for groups 1, 2, and 3 for the continuous time model.}
\label{fig:Cont}
\end{figure}

\begin{figure}[!htb]
\centering
\includegraphics[width=0.45\textwidth]{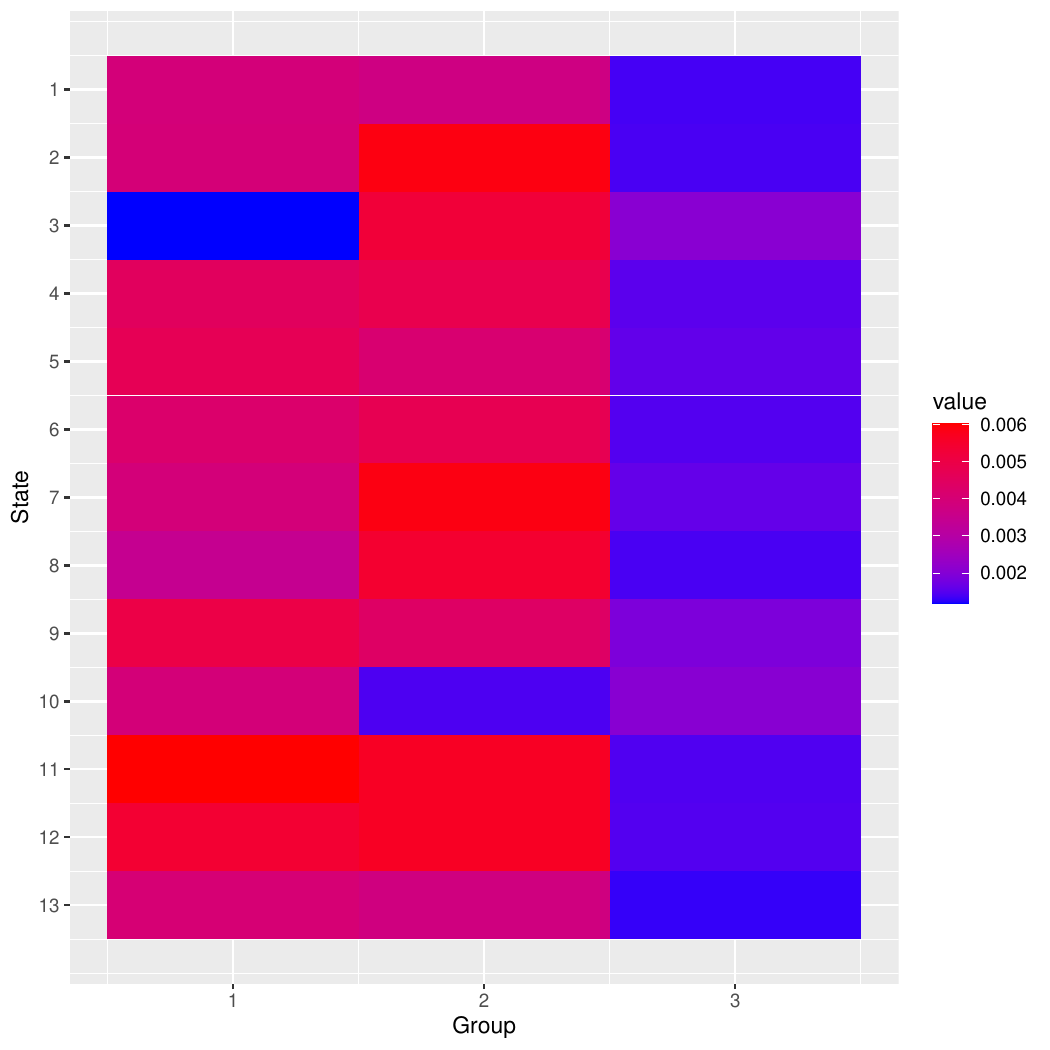}
\caption{Heat map of estimated rates for groups 1, 2, and 3 for the continuous time model.}
\label{fig:conttimes}
\end{figure}

In Figure \ref{fig:Cont}, heat maps of the transition matrices taken from the estimated infinitesimal generators are shown.  Additionally, in Figure \ref{fig:conttimes}, a heat map of the estimated exponential rates is shown. Regarding the transition probabilities, we again see users in all groups having a fairly high probability of transitioning to State 13 from all states; however, there are some exceptions. For example in Group 1, users in States 7 or 10 have a very low probability of transitioning to State 13. 

The estimated rates reveal some interesting characteristics of each group. Group 3 is characterized by spending a longer time on average in all states compared to Groups 1 and~2, with possibly a slightly shorter period of time on average in States 3, 9, and 10. In Group 1, users appear to spend a short period of time on average in all states, with a longer amount of time on average in State 3, and a particularly short period of time in States 9, 11, and 12 . In group 2, users spend a longer period of time on average in State 10 compared to the other states. Although this application is in the area of online marketing, the results for Groups 1 and 2 could be of particular interest in the area of criminology if States 3 or 10 were to contain inappropriate subject matter.

Finally, a classification table comparing the classifications given by the discrete and continuous time models, respectively, is shown in Table \ref{tab:class}. It is clear that although both models found the same number of clusters, the classifications are very different.
\begin{table}[h]
\centering
\caption{Classification table comparing the clusters found by the discrete and continuous time models for the YouChoose dataset.}
\begin{tabular}{c|ccc}
\hline
&\multicolumn{3}{|c}{Continuous}\\
\hline
Discrete&1&2&3\\
1&1636& 1122& 1311\\
  2 & 1440& 3797& 143\\
  3& 2034&  340& 2738\\
\hline
  \end{tabular}
  \label{tab:class}
\end{table}
\section{Summary}\label{dis}

An approach was introduced that incorporates continuous time for unsupervised and semi-supervised classification of clickstream data. This approach is based on a mixture of first-order continuous time Markov models. An EM algorithm was outlined for parameter estimation, and the BIC was used to select the number of groups~$G$.

In the analyses that were carried out, we noted that incorporating the amount of time spent in each category allowed for the detection of groups of users that the discrete time model was unable to detect. This was especially true where there was not a lot of separation in the transition probabilities between groups, but differences in the average amount of time spent in each state. Moreover, if the amount of time spent in each state was similar, on average, between groups but there was a lot of separation in the transition probabilities, the continuous time model performance was very similar to the discrete time model. Finally, when applied to the YouChoose data, although the continuous and discrete time models found the same number of groups, they did not result in the same classification. Furthermore, the continuous time model provided some valuable insight into the behaviour of the users. Specifically, users in Group 1 spent longer in state 3 than the other states, and users in Group 2 generally spent longer in state 10. Although this example is in the area of online marketing, these results may indicate that the continuous time model could prove useful in many other areas.

In future work, a different distribution for the holding time could be considered. Although the classical approach is to use an exponential holding time in each state, this may not be realistic in some real applications. The issue with the exponential distribution is that the model allows for almost immediate or unrealistically long transition times. This leads us to another issue, i.e., that the continuous time model is time unit dependent, which also needs to be considered in a real application. Therefore, the of use a truncated exponential distribution for the holding time will be a topic of future work. Finally, although conceived in the context of clickstream data, this methodology could be used in other applications that look at state transitions, such life events, illnesses, and migration patterns.

\section*{Conflict of Interest Statement}

The authors declare that the research was conducted in the absence of any commercial or financial relationships that could be construed as a potential conflict of interest.

\section*{Author Contributions}

MG and PM conceived and developed the idea. MG led the model development and carried out the implementation. PM advised on the model development. MG carried out the analyses. MG and PM wrote the manuscript. 

\section*{Funding} 

This work was supported by a Vanier Canada Graduate Scholarship and Banting Postdoctoral Fellowship from the Natural Sciences and Engineering Research Council of Canada (MG), the Canada Research Chairs program (PM), and an E.W.R.~Steacie Memorial Fellowship (PM).

\end{document}